\documentclass[prb,superscriptaddress,floatfix,letterpaper,twocolumn,aps,showpacs]{revtex4-1}

\usepackage[utf8]{inputenc}
\usepackage[usenames,dvipsnames,svgnames,table]{xcolor}
\usepackage{natbib,amsmath,amsbsy,amssymb,scrextend,graphicx,color,placeins,array,siunitx}
\setcitestyle{square, comma, numbers}
\usepackage{hyperref}

\renewcommand{\th}{\mathrm{th}}
\renewcommand{\cite}[1]{[\onlinecite{#1}]}

\newcommand{\ff}{\mathrm{ff}}
\newcommand{\dep}{\mathrm{dp}}
\newcommand{\trap}{\mathrm{trap}}
\newcommand{\pin}{\mathrm{pin}}
\newcommand{\C}{\bar{C}}
\renewcommand{\ff}{\mathrm{ff}}

\begin{document}

%\title{Quantitative study of thermal creep near depinning in type-II
%superconductors} 
%
%\title{Thermal creep in type-II superconducting films analyzed within the
%strong pinning paradigm}
%
%\title{Experimental test of thermal creep within the strong pinning paradigm}
%
\title{Experimental test of strong pinning and creep in current--voltage
characteristics\break of type II superconductors}

\author{M.\ Buchacek}
\affiliation{Institute for Theoretical Physics, ETH Zurich, 8093 Zurich, Switzerland}
\author{Z.\ L.\ Xiao}
\affiliation{Materials Science Division, Argonne National Laboratory, 
Lemont, IL 60439, USA}
\author{S.\ Dutta}
\affiliation{Tata Institute of Fundamental Research, Homi Bhabha Road, Colaba, Mumbai 400005, India}
\author{E.\ Y.\ Andrei}
\affiliation{Department of Physics and Astronomy, Rutgers University, Piscataway, New Jersey 08855, USA}
\author{P.\ Raychaudhuri}
\affiliation{Tata Institute of Fundamental Research, Homi Bhabha Road, Colaba, Mumbai 400005, India}
\author{V.B.\ Geshkenbein}
\affiliation{Institute for Theoretical Physics, ETH Zurich, 8093 Zurich, Switzerland}
\author{G.\ Blatter}
\affiliation{Institute for Theoretical Physics, ETH Zurich, 8093 Zurich, Switzerland}

\date{\today}

\begin{abstract}
Pinning and creep determine the current--voltage characteristic of a type II
superconductor and thereby its potential for technological applications.  The
recent development of strong pinning theory provides us with a tool to assess
a superconductor's electric properties in a quantitative way.  Motivated by
the observation of typical excess-current characteristics and field-scaling of
critical currents, here, we analyze current--voltage characteristics measured
on 2H-NbSe$_2$ and $a$-MoGe type II superconductors within the setting
provided by strong pinning theory. The experimentally observed shift and
rounding of the voltage-onset is consistent with the predictions of strong
pinning in the presence of thermal fluctuations. We find the underlying
parameters determining pinning and creep and discuss their consistency.
\end{abstract}

\maketitle 

\section{Introduction}

\begin{figure}[ht]
    \centering
	\includegraphics[scale=1]{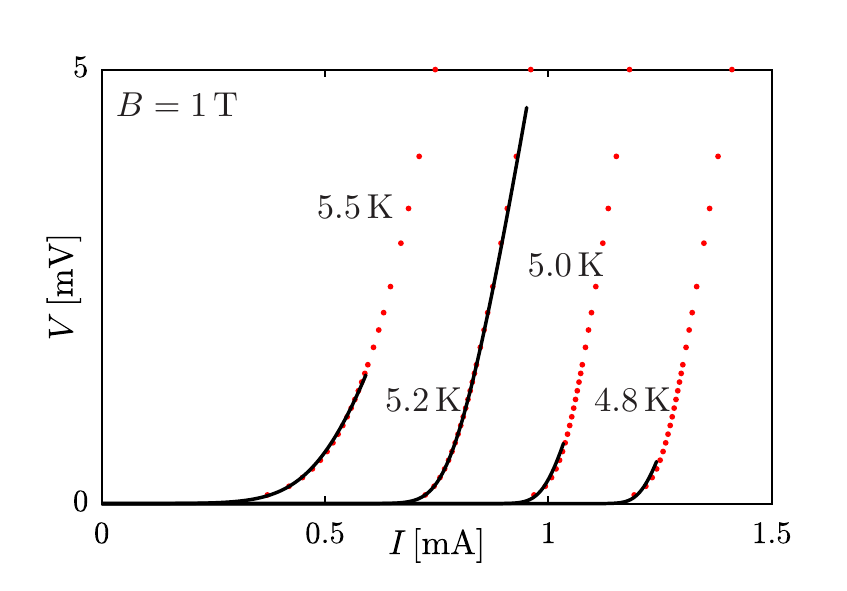}\\
	\caption{Current--voltage characteristic measured on a 2H-NbSe$_2$
	sample at a fixed field $H = \SI{1.0}{T}$ and temperatures $T =
	4.8,\,5.0,\,5.2$, and $\SI{5.5}{K}$ (red points).  Black lines are
	fits to the data based on the prediction of strong pinning theory;
	fits are restricted to the region of applicability of the theory (see
	text for details). With increasing temperatures, the characteristic
	shifts to the left and the rounding in the critical region at voltage
	onset becomes more pronounced.  } \label{Fig:Andrei_data}
\end{figure}

\begin{figure}[ht]
    \centering
        \includegraphics[scale = 1]{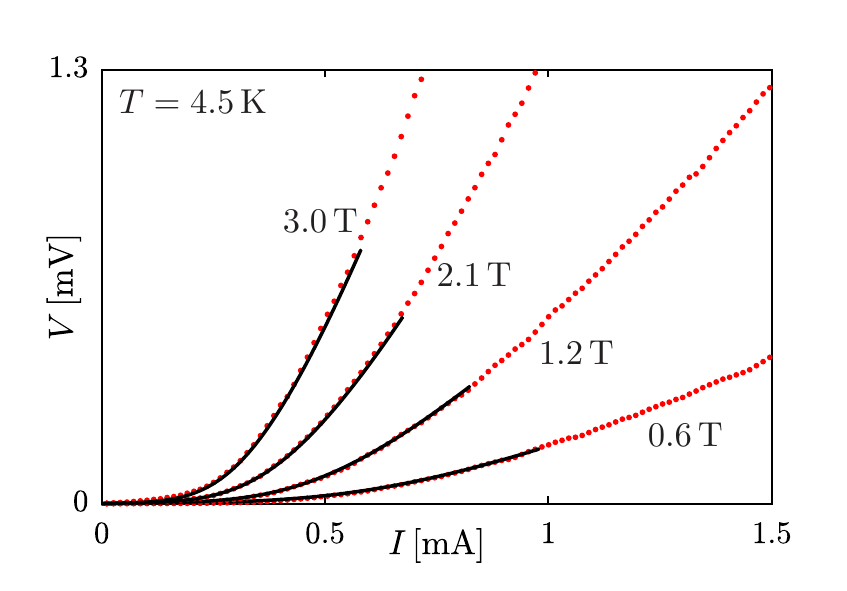}\\
	\caption{Current--voltage characteristic measured on a $a$-MoGe sample
	at fixed temperature $T = \SI{4.5}{K}$ and applied fields $H = 0.6,
	1.2,2.1$ and $\SI{3.0}{T}$ (red points).  Black lines are fits to the
	data based on the prediction of strong pinning theory; fits are
	restricted to the region of applicability of the theory (see text for
	details). With increasing fields, the characteristic shifts to the
	left and the slope of the flux-flow response at large currents
	increases.  } \label{Fig:Pratap_data}
\end{figure}

Topological excitations appearing in the ordered phase of many materials have
a strong impact on their physical properties. Such excitations interact with
material's defects, what modifies both the structural and dynamical properties
of the topological superstructure and of the host material itself.  In type-II
superconductors, the topological objects appear in the form of vortices due to
(current-)induced or applied magnetic fields \cite{Abrikosov1957}.  While free
moving vortices result in a finite resistivity \cite{Bardeen1965}, pinning the
vortices to material defects \cite{Anderson1962} helps maintaining the
superconductor's dissipation-free transport of electric current.  In the
absence of fluctuations, vortex motion only appears upon exceeding the
critical current $I_c$.  Thermal fluctuations potentially modify this picture
by allowing for a slow, creep-type vortex motion even at subcritical drives $I
< I_c$ that leads to a shift and smoothing of the transition in the critical
region.  In this paper, we make use of the quantitative results from strong
pinning theory \cite{Buchacek2018, Buchacek2019_theory} in order to
unambiguously identify vortex-creep in the critical region of the
current--voltage characteristic of two distinct low-$T_c$ materials,
2H-NbSe$_2$, with $T_c = \SI{7.18}{K}$ \cite{Xiao2002}, see Fig.\
\ref{Fig:Andrei_data}, and \textit{a}-MoGe, with $T_c = \SI{6.9}{K}$
\cite{Roy2019}, see Fig.\ \ref{Fig:Pratap_data}.

The origins of pinning and creep can be traced back to the seminal papers of
Anderson, Kim, and collaborators \cite{Anderson1962, AndersonKim1964,
KimHempStrnad1963}, where Abrikosov vortices pinned onto defects were taken
responsible for the properties of current transport in hard superconductors.
Besides setting a focus on the Bean critical state \cite{Bean1962} and its
log-time decay, the shape of the current--voltage characteristic in the
critical region was discussed as well, including an interpolation formula
describing the transformation of creep-type to flow-type response of vortices
that prevail at low and high drives, respectively \cite{AndersonKim1964}.
Later, much further work has been devoted to studying creep, particularly in
the high-$T_c$ superconductors where thermal fluctuations play an important
role. On the one hand, relaxation experiments at low drives helped to identify
glassy physics characterized by diverging barriers \cite{Yeshurun1996}, while
resistive measurements using sensitive voltmeters served the similar purpose
of identifying a non-linear, i.e., glassy response \cite{Safar1992}. However,
less attention was given to the behavior in the critical region, e.g., the
vanishing of barriers near $j_c$ or the smoothing of the characteristic. Not
least, this is due to the inadequacy of pinning theories to make quantitative
predictions, a deficiency that was overcome only recently, at least for the
case of strong pinning \cite{Buchacek2018, Buchacek2019_theory}.

A distinctive feature of strong pinning is its excess-current characteristic, an $I-V$
characteristic that exhibits a linear (flux-flow) response at large drives $I
> I_c$ that is shifted by the critical current $I_c$
\cite{Thomann2012,Thomann2017}, in an idealized $T=0$ situation $V =
R_\mathrm{ff}(I-I_c) \Theta(I-I_c)$ with $R_\mathrm{ff}$ the flux-flow
resistance. Apart from the datasets \cite{Xiao2002,Roy2019} analyzed in great
detail below, such characteristics have been observed in other recent
\cite{Sacepe2018} as well as older, even textbook \cite{Strnad1964}
experiments. The analysis and proper understanding of changes in this $I$--$V$
characteristic with increasing temperature $T$ is the central topic of this
work and involves the following goals: i) demonstrate the power and
consistency of strong pinning theory in quantitatively explaining experimental
data of $I$--$V$ characteristics in different materials for different
temperatures $T$ and magnetic fields $B$.  ii) Extract fitting parameters and
check for their qualitative consistency with expectations from vortex theory
as obtained within the Ginzburg-Landau (GL) phenomenological framework.

In pursuing this program, we have to disentangle two effects of temperature
$T$, one being encoded in the parameter $\tau = 1-t = 1- T/T_c$ that appears
in the Ginzburg-Landau (GL) mean-field theory of the superconducting state,
the other being the temperature $T$ as the driver of thermal fluctuations.
While the former lives on the scale $T_c$, the scale of the latter is given by
the fluctuation energy $\xi\varepsilon_0$, with $\varepsilon_0 = (\Phi_0/4 \pi
\lambda)^2$ the vortex line energy, $\Phi_0 = hc/2e$ is the magnetic flux
quantum, and $\xi$, $\lambda$ denote the correlation- and screening lengths of
the superconductor.  A further dependence is due to the presence of the
magnetic field (or more precisely, induction) $B$, introducing the distance
$\tau_b = 1 - T/T_c - B/H_{c2}(0) = 1 - t - b$ from the upper critical
field-line $H_{c2}(T)$.

The critical current $I_c$ and the flux-flow resistance $R_\mathrm{ff}$, the
parameters quantifying the shape of the excess-current characteristic, depend
on the temperature $T$ and magnetic field $B$ via $\tau$ and $b$ (or
$\tau_b$); approaching the $H_{c2}(T)$-line in the $H$--$T$ phase diagram,
$I_c(B,T)$ decreases, see Fig.\ \ref{Fig:Andrei_data}, and the flux-flow
becomes steeper, see Fig.\ \ref{Fig:Pratap_data}, with $R_\mathrm{ff}$
approaching the normal-state resistance $R_n$. Although these GL predictions
are in rough agreement with experimental data, they do not catch effects of
thermal fluctuations that are manifest in the data as well.  In particular,
within strong pinning theory, the inclusion of fluctuations predicts a further
downward shift in the excess current, replacing the critical current $I_c$ by
the depinning current $I_\dep(T)$, and a rounding of the transition to the
ohmic branch of the characteristic \cite{Buchacek2018, Buchacek2019_theory}.

A consistent analysis of creep phenomena then requires to separate these
different types of temperature dependence in the experimental data. In our
analysis, we achieve this task by rescaling the data to make it collapse onto
one curve at asymptotically large drives. The comparison of the
temperature-dependent rounding and shift of the collapsed data in the critical
region around $I_\dep$ with the prediction from strong pinning theory then
provides clear evidence for vortex creep, see Figs.~\ref{Fig:fits_Andrei} and
\ref{Fig:fits_Pratap}. In figures \ref{Fig:Andrei_data} and
\ref{Fig:Pratap_data}, we show the corresponding fits to the original
experimental data that demonstrate an impressive agreement.  Furthermore, our
quantitative theory allows to extract important parameters of vortex physics
from the data and check for their internal consistency. Note the difference to
the celebrated theory of weak collective pinning
\cite{Larkin1979,Blatter1994}, where the addition of forces due competing
defects poses a formidable task; the latter is straightforward within the
strong pinning paradigm where the density $n_p$ of defects is assumed to be
small such that pins act individually. As a consequence, results obtained
within strong pinning theory can be pushed to provide a numerical accuracy
beyond what can be achieved within the framework of weak collective pinning.
In particular, strong pinning theory can offer quantitative expressions for
critical current densities \cite{IvlevOvchinnikov91}, current--voltage
characteristics \cite{Thomann2012,Thomann2017}, and thermal creep in the
critical regime \cite{Buchacek2018, Buchacek2019_theory}.

In the following, we first introduce and discuss the result for the
current--voltage characteristic as obtained from strong pinning theory
\cite{Buchacek2018,Buchacek2019_theory}, see Sec.\ \ref{sec:spt}. In Sec.\
\ref{sec:exp}, we present the experimental data on the current--voltage
characteristics of 2H-NbSe$_2$ and $a$-MoGe and extract the parameters of the
characteristic.  Sec.\ \ref{sec:barriers} is devoted to the analysis of creep
barriers, where we put forward a new type of analysis that aims at
$U(F_\pin)$, i.e., the barrier's dependence on the pinning-force density
$F_\pin$ rather than the usual dependence $U(F_{\rm \scriptscriptstyle L})$
involving the Lorentz-force density $F_{\rm \scriptscriptstyle L}$ that drives
the vortices. In Sec.\ \ref{sec:par}, we relate the parameters in the
current--voltage characteristic as obtained from the comparison with
experimental data to the `microscopic' parameters of strong pinning theory; we
summarize our results in Sec.\ \ref{sec:sc} and provide some concluding
remarks.

\section{Current--Voltage Characteristic}\label{sec:spt}
We discuss the excess-current characteristic in the presence of thermal
fluctuations (creep) as derived within the strong pinning paradigm.  Within
standard vortex physics, the relation between the driving current density $j$
and the vortex velocity $v$ is obtained from the dissipative equation of
motion balancing the effects of the current-induced Lorentz force density
$F_{\rm \scriptscriptstyle L}(j) = jB/c$ driving the vortices, the pinning
force density $F_\pin(v,T)$ due to the defects, and the viscous force density
$-\eta v$ proportional to the vortex velocity $v$,
\begin{align}\label{Eq:force_balance}
   \eta v = F_{\rm \scriptscriptstyle L}(j) - F_\pin(v,T).
\end{align}
In the absence of thermal creep ($T = 0$) and at low velocities, strong
pinning theory predicts a nearly constant pinning-force density $F_\pin\approx
F_c$, $F_c$ the critical force density, over a large regime of velocities
\cite{Thomann2012,Thomann2017}, in agreement with Coulomb's law of friction.
For small currents, the driving Lorentz force then can be compensated by the
pinning force and vortices remain pinned, $v=0$.  A finite vortex velocity $v
\approx (F_{\rm \scriptscriptstyle L}(j)-F_c)/\eta >0$ only appears at larger
drives; as a result, we find the excess-current characteristic with vanishing
voltage below the critical current density $j_c = cF_c/B$ and a shifted ohmic
branch above.  This seemingly trivial result owes its validity to the
separation of the two velocity scales $v_c \ll v_p$ describing the average
motion $v_c = F_c/\eta$ of the vortex lattice and the velocity $v_p \sim
f_p/\eta a_0^3$ of vortices during individual (de-)pinning events (with $f_p$ the
pinning force of an individual defect and $a_0 = \sqrt{\Phi_0/B}$ the separation between vortices); as shown in Refs.\
\cite{Thomann2012,Thomann2017}, the pinning-force density $F_\pin(v,T=0)$
changes with velocity on the scale $v_p$.

At finite temperature $T > 0$, thermal creep facilitates the escape of
vortices from pinning defects; such creep motion is characterized by an energy
barrier $U(v)$ which relates to the velocity $v$ of vortices via
\begin{align} \label{Eq:U_T-v}
   U(v) \approx k_{\rm \scriptscriptstyle B} T \ln (v_\th/v).
\end{align}
Here, $v_\mathrm{th}$ is the thermal velocity scale, related to an attempt
frequency for thermal depinning and derived within strong pinning theory in
Refs.\ [\onlinecite{Buchacek2018, Buchacek2019_theory}], see also Sec.\
\ref{sec:par}. On approaching the thermal velocity scale $v_\th$, vortices
traverse the pins sufficiently fast and the barrier slowing down the motion
becomes irrelevant.  The second central result provided by strong pinning
theory is the force-dependence of these very same barriers, which assumes the
simple form
\begin{align} \label{Eq:U_F}
   U[F_\pin(v,T)] \approx U_c\,[1- F_\pin(v,T)/F_c]^{3/2}.
\end{align}
This result involves two noteworthy features: first, the relevant force in
this simple relation is not the usual driving Lorentz-force density $F_{\rm
\scriptscriptstyle L} \propto j$ (that would result in a standard relation
\cite{Blatter1994} $U(j)$) but it is the pinning-force density $F_\pin(v,T)$.
Second, the exponent 3/2 is universal for any smooth pinning potential; its
origin is found \cite{Buchacek2019_theory} in the thermally induced shift
$\delta x$ of the (de-)pinning point which relates to the barrier $U$ via the
scaling $U \propto \delta x^{3/2}$. A detailed derivation of $U_c$ is given in
Refs.\ [\onlinecite{Buchacek2018, Buchacek2019_theory}], see also Sec.\
\ref{sec:par}.  The two equations \eqref{Eq:U_T-v} and \eqref{Eq:U_F} combine
into a velocity and temperature dependence of the pinning force density
$F_\pin(v,T)$ in the form
\begin{align} \label{Eq:F_v-T}
   F_\pin(v,T)/F_c \approx  1-[(k_{\rm \scriptscriptstyle B} T/U_c)
   \log (v/v_\th)]^{2/3}.
\end{align}
Inserting the expression for $F_\pin(v,T)$ into the equation of motion
Eq.~\eqref{Eq:force_balance} and dividing by $F_c$, we arrive at a simple
formula \cite{Buchacek2018, Buchacek2019_theory} for the fluctuation-enhanced
vortex velocity or current--voltage characteristic,
\begin{align} \label{Eq:j-v}
   {v}/{v_c} = {j}/{j_c} - 1
   + \bigl[({k_{\rm \scriptscriptstyle B} T}/{U_c})
   \ln ({v_\th}/{v})\bigr]^{2/3}.
\end{align}
Here, we have used the definition of the free flux-flow velocity $v_c =
F_c/\eta$ at $F_c$. The dynamical equation \eqref{Eq:j-v} captures the small
vortex velocity at subcritical drives $j < j_c$, the rounding of the
characteristic in the critical region, and the (initial part) of the smooth
approach to the ohmic region. As $v$ approaches $v_\mathrm{th}$, thermal
fluctuations become irrelevant and the characteristic joins the excess-current
shape.

The calculation leading to Eq.\ \eqref{Eq:j-v} is based on Kramers' rate
theory \cite{Kramers1940,Hanggi1990} assuming an activation barrier $U_c \gg
k_{\rm \scriptscriptstyle B} T$.  Strong pinning theory tells
\cite{Buchacek2018}, that the relevant barrier depends on velocity via $U
\approx k_{\rm \scriptscriptstyle B} T \log (v_\th/v)$, see Eq.\
\eqref{Eq:U_T-v}, that restricts the applicability of Eq.~\eqref{Eq:j-v} to
$v\lesssim v_\th/e$, see Figs.~\ref{Fig:Andrei_data} and \ref{Fig:Pratap_data}.

We relate the theoretical result \eqref{Eq:j-v} to the experimentally
accessible current $I = A j$ and voltage $V = LE$ ($E = Bv/c$ the electric
field) using the sample geometry (length $L$ and area $A$) and the definition
of the free flux-flow voltage at $I_c$, $V_c = R_\mathrm{ff} I_c$,
\begin{align} \label{Eq:I-V}
   V = R_\mathrm{ff}(I - I_c) + V_c
   \Bigl[\frac{k_{\rm \scriptscriptstyle B} T}{U_c}
   \log\Bigl(\frac{v_\th}{v_c} \frac{V_c}{V}\Bigr)\Bigr]^{2/3}.
\end{align}
This result can be directly compared to the data; it involves the four
parameters $I_c, R_\mathrm{ff}, U_c/k_{\rm \scriptscriptstyle B} T, v_\th/v_c$
that are obtained in two steps. At large voltages $V$ (or velocities $v$),
creep is irrelevant and the characteristic reduces to the simple
exccess-current form $V \approx R_\mathrm{ff}(I-I_c)$, from which $I_c$ and
$R_\mathrm{ff}$ can be directly read off. In a next step, we rescale the data
$V \to V/V_c = v/v_c$ and $I \to I/I_c = j/j_c$ to bring it to the form of
Eq.\ \eqref{Eq:j-v}. This rescaling induces a data collapse in the asymptotic
region; the deviations appearing in the transition region then are due to
creep. It is this deviation from which we can extract the two remaining
parameters $U_c/k_{\rm \scriptscriptstyle B} T$ and $v_\th/v_c$ through a
careful fit to the experimental data in the transition region.

\section{Experimental Current-Voltage Characteristics}\label{sec:exp}

We illustrate the procedure outlined above for fitting the data and extracting
the relevant physical parameters $j_c, v_c, U_c, v_\th$ for the two
low-$T_c$ superconductors 2H-NbSe$_2$ and $a$-MoGe.

\subsection{2H-NbSe$_2$}

\begin{figure}[hbt]
    \centering
        \includegraphics[scale=0.9]{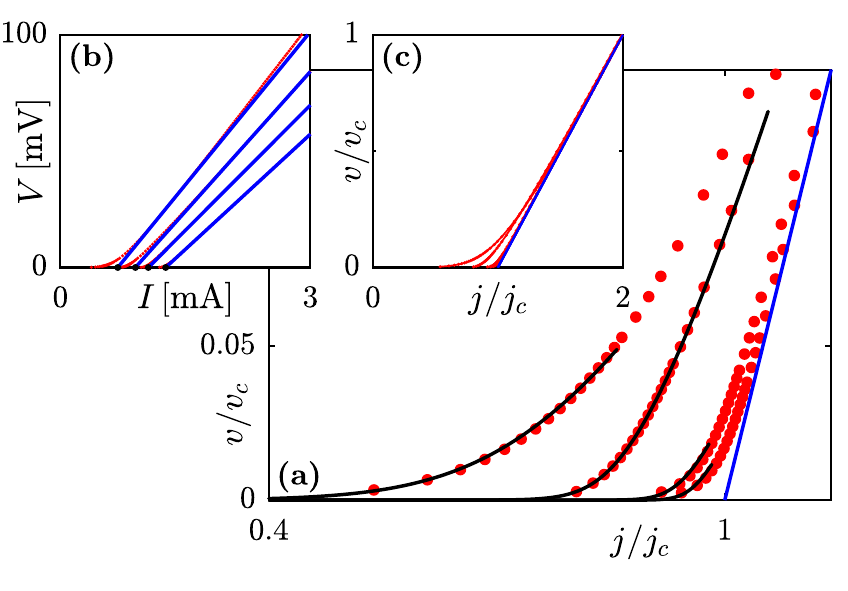}\\
	\caption{Observation of thermal creep in 2H-NbSe$_2$. (a): evolution
	of excess-current characteristic with increasing temperature $T$: red
	data points are taken at $B = \SI{1.0}{T}$, ${T = 4.8,\,5.0,\,5.2}$ and $\SI{5.5}{K}$,
	black lines are fits to Eq.~\eqref{Eq:j-v} describing the creep
	characteristic within strong pinning theory. Large temperatures
	produce a marked rounding of the characteristic in the critical region
	near voltage onset and the $T = 0$ excess-current characteristic
	(solid blue) is approached at larger drives.  Fits are shown up to
	velocities where activation barriers remain larger than temperature
	$T$.  Inset (b) shows the raw experimental data, see Fig.\
	\ref{Fig:Andrei_data}, while the inset (c) presents the data collapsed
	to a single curve at large drives; such scaling provides the
	parameters $j_c$ and $v_c$.}
	\label{Fig:fits_Andrei}
\end{figure}

The measurements were performed in a $H = 1$ T field directed along the
$c$-axis with an in-plane dc current applied through the cross-sectional area
$A = d\, w$, $d = \SI{0.02}{mm}$, $w = \SI{0.47}{mm}$ in a sample of length $L
= \SI{8}{mm}$. The inter-vortex distance $a_0 = \sqrt{\Phi_0/B}\approx
\SI{0.43}{nm}$ is small compared to the sample thickness $d$ and hence the
standard strong pinning theory for 3D bulk pinning \cite{Blatter2008} is
applicable.  Fig.~\ref{Fig:fits_Andrei} shows the data and fitting to Eq.\
\eqref{Eq:j-v} in the critical region, with the fits restricted to the region
$v\lesssim v_\th/{\rm e}$ where our theory applies.

\begin{figure}[ht]
    \centering
        \includegraphics[scale=0.9]{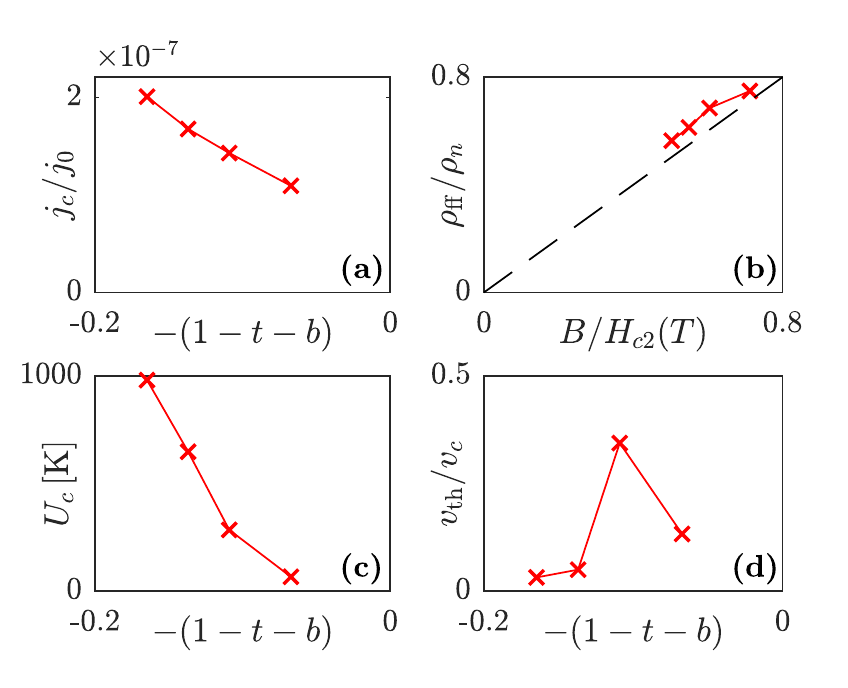}\\
	\caption{Parameters for 2H-NbSe$_2$ extracted from fitting the
	$I$--$V$ characteristics of Fig.\ \ref{Fig:fits_Andrei} as a function
	of $-\tau_b = -(1 - t - b)$ chosen such as to approach the $H_{c2}$ line
	from the left (as in the standard $H$--$T$ diagram). The small value
	of the scaling parameter $j_c/j_0$ in (a) testifies for a small defect
	density $n_p$.  The flux-flow resistivity $\rho_\ff/\rho_n$ shown in
	(b) follows the Bardeen-Stephen law. The activation barrier $U_c$
	decays on approaching the upper-critical field line, see (c), while
	the scaled thermal velocity $v_\mathrm{th}/v_c$ shown in (d) first
	rises on approaching the $H_{c2}$-line and then drops, that can be
	consistently explained by strong pinning theory, see Sec.\
	\ref{sec:par}.} \label{Fig:parameters_Andrei}
\end{figure}

At high velocities $v > v_\th$, creep is irrelevant and we fit the data to the
excess-current characteristic $V = R_\mathrm{ff}(I - I_c)$, see blue lines in
Fig.\ \ref{Fig:fits_Andrei}(b). Analyzing the four curves at $T =
4.8,\,5.0,\,5.2$, and $\SI{5.5}{K}$, we obtain the critical current densities
$j_c = 13,\,11,\,9.6,\,$ and $\SI{7.3}{A/cm^2}$. These values are far below
the ($T=0$) depairing current density $j_0\approx \SI{6.7e7}{A/cm^2}$,
consistent with a small defect density $n_p$, see Sec.\ \ref{sec:par}, and
decrease on approaching the $H_{c2}$-line, $\tau_b = 1-t-b \to 0$, see Fig.\
\ref{Fig:parameters_Andrei}(a). From the slopes, we obtain the flux-flow
resistivities $\rho_\mathrm{ff}$ and using the normal state resistivity
$\rho_n\approx \SI{6.9e-3}{\ohm cm}$, we verify the consistency with the
Bardeen-Stephen result $\rho_\ff/\rho_n\approx B/H_{c2}$, see Fig.\
\ref{Fig:parameters_Andrei}(b). The vortex motion generates the electric field
$V/L = E = Bv/c$ and we obtain the estimates for the free flux-flow velocities
$v_c = c V_c/BL = (5.2,\,4.7,\,4.5,\,3.8)\times 10^2\,\mathrm{cm}/\mathrm{s}$.

While the original data include the intrinsic field- and temperature
dependences of $j_c$ together with the rounding in the critical region, the
rescaled data $V/V_c = v/v_c$ and $I/I_c = j/j_c$ in Fig.\
\ref{Fig:fits_Andrei}(c) collapse at high velocities to a single line of unit
slope; the temperature-dependent rounding and shift of the curves away from
the excess-current characteristic in the critical region then can be firmly
attributed to thermal creep and serves to find the remaining parameters
$U_c/k_{\rm \scriptscriptstyle B} T$ and $v_\th/v_c$.

Let us then focus on the most interesting part of the characteristic, the
smooth transition to flux-flow in the critical region around $j_c$, see Fig.\
\ref{Fig:fits_Andrei}(a). In Sec.\ \ref{sec:barriers} below, we present a
protocol for the optimal extraction of these parameters by replotting the
current-voltage data in a form that accounts for the creep-type motion in this
regime. Inserting the results back into the characteristic \eqref{Eq:j-v}, we
obtain excellent fits to the data; the extracted barrier $U_c$ of order 1000 K
rapidly decreases when approaching the $H_{c2}$-line, $\tau_b = 1-t-b \to 0$,
see Fig.\ \ref{Fig:parameters_Andrei}(c), in agreement with strong pinning
theory and further discussion in Sec.\ \ref{sec:par}.  Finally, the results
for the thermal velocity parameter $v_\th/v_c$ are shown in Fig.\
\ref{Fig:parameters_Andrei}(d). The discussion in Sec.\ \ref{sec:par} predicts
an increase of $v_\th/v_c$ with temperature that is consistent with the
experimental findings; the drop near the phase boundary may be due to a
collapse of strong pinning $\kappa \to 1$.  Its numerical value turns out
about an order of magnitude larger than expected from strong pinning theory,
however, we note that we have least control on this quantity since it assumes
the role of an attempt frequency in Kramer's rate theory, a quantity that is
notoriously difficult to calculate.

\subsection{\textit{a}-MoGe}

In a similar fashion, we analyse the \textit{I-V} measurements on
\textit{a}-MoGe films reported in Ref.\ [\onlinecite{Roy2019}], with data
available both at different fields and temperatures. The applied magnetic
field in the range $H = 0.03-\SI{7} {T}$ (the upper-critical field is
${H_{c2}(0) = \SI{13}{T}}$) implies a vortex lattice constant $a_0 =
17-\SI{262}{nm}$. The current $I$ is applied along the direction of the film
of length $L = \SI{1}{mm}$. The thickness iof the film measures $d =
\SI{20}{nm}$ and its width is $w = \SI{300}{\mu m}$; while $d > a_0$ above $H
\sim \SI{4} {T}$, the low-field region may crossover to 2D pinning, see
further discussion below.
\begin{figure}[h]
    \centering
        \includegraphics[scale=0.9]{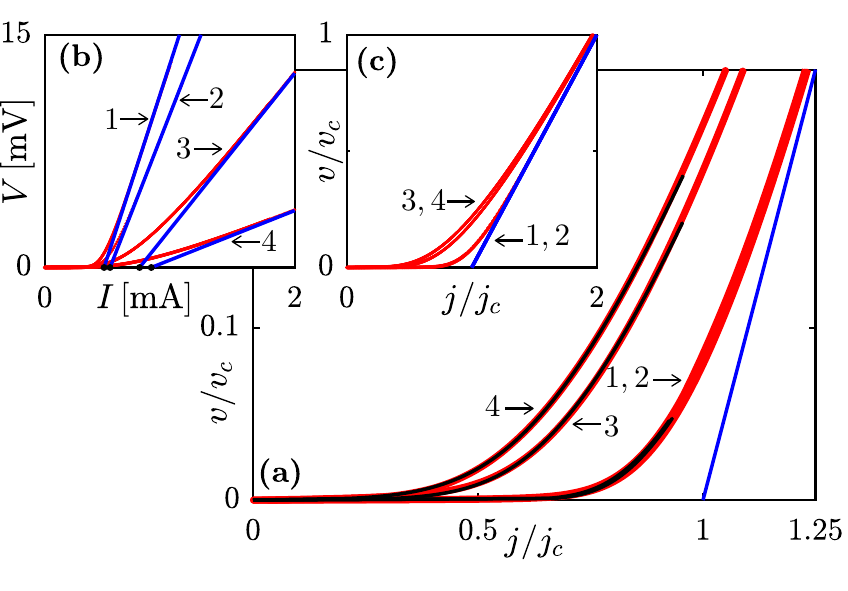}
	\caption{(a) Extrapolated $T=0$ excess-current characteristic (blue)
	and experimental data (red) at finite temperatures 1:
	$\SI{0.28}{K},\,\SI{1.2}{T}$, 2: $\SI{0.45}{K},\,\SI{1.0}{T}$ (these
	two datasets collapse to an almost identical curve), 3:
	$\SI{2.0}{K},\,\SI{0.5}{T}$, and 4: $\SI{3.5}{K},\,\SI{0.2}{T}$.
	Thermal fluctuations produce a dowmward shift and rounding of the
	characteristic in the critical regime. Black lines provide excellent
	fits within strong pinning theory; the fit stops when barriers $U_c$
	approach $k_{\rm \scriptscriptstyle B} T$. The insets (b) and (c) show
	the raw and rescaled experimental data.} 
	\label{Fig:fits_Pratap}
\end{figure}

\begin{figure}[ht]
    \centering
        \includegraphics[scale=0.9]{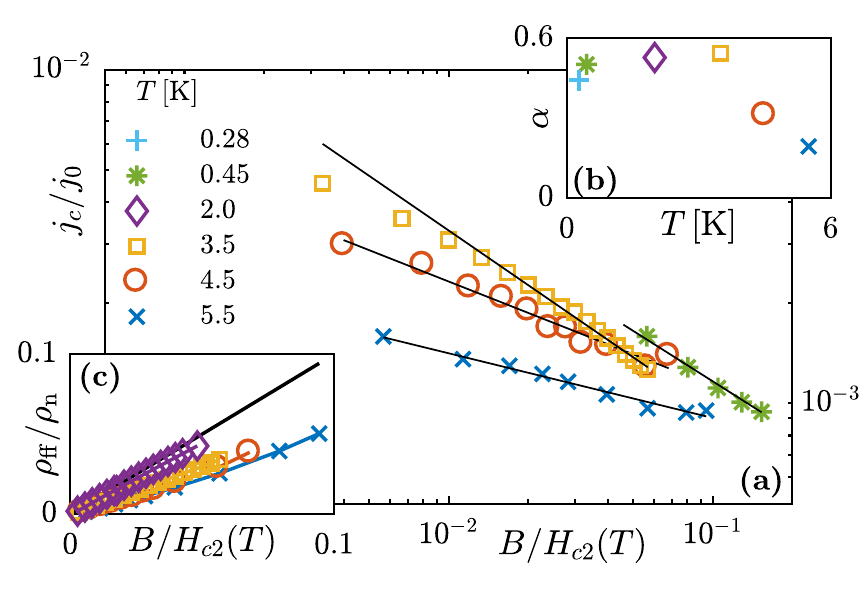}
	\caption{(a) Critical current densities (a) versus magnetic field
	fitted to a power law $j_c\propto B^{-\alpha}$, with the exponents
	$\alpha$ at different temperatures shown in (b). The exponent $\alpha
	\approx 0.5$ is in good agreement with the prediction from strong
	pinning theory, and its decrease at high temperatures matches recent
	numerical results \cite{Willa2017}.  (c) Flux-flow resistivity
	compared to the Bardeen-Stephen formula (solid line); values below the
	Bardeen-Stephen line are consistent with more detailed predictions of
	Larkin and Ovchinikov \cite{Larkin1986}.} \label{Fig:Pratap_j_c}
\end{figure}

In Fig.\ \ref{Fig:fits_Pratap}, we analyze several $I-V$ curves taken at
finite temperatures $T = 0.28,\,0.45,\,2$, and $\SI{3.5}{K}$. The results of
the scaling collapse of this data, providing the parameters $j_c$ and
$\rho_\mathrm{ff}$, are shown in Fig.\ \ref{Fig:Pratap_j_c}. Making use of
field-dependent data, we find that the critical current density scales as $j_c
\sim B^{-\alpha}$ (Fig.~\ref{Fig:Pratap_j_c}(a), with $j_0\approx
\SI{6.8e6}{A\per cm^2}$) with the exponent $\alpha \approx 0.6$ measured at
low temperatures, consistent with theoretical predictions for the strong
pinning scenario \cite{IvlevOvchinnikov91, Thomann2017, Kwok2016}.  The
exponent decreases towards $\alpha\approx 0.3$ at higher temperatures, in
agreement with numerical simulations \cite{Willa2017} reporting such a
behavior with increasing vortex core size. At low fields, a crossover to 2D or
1D strong pinning may occur, see discussion in Sec.\ \ref{sec:par} below. The
resistivity extracted from the flux-flow regime above $j_c$ remains below the
Bardeen-Stephen estimate (see Fig.\ \ref{Fig:Pratap_j_c}(d), $\rho_n \approx
\SI{1.57e-4}{\ohm\cm}$), that is qualitatively consistent with a more
elaborate result of Larkin and Ovchinikov \cite{Larkin1986}.  The flux-flow
vortex velocities corresponding to the analyzed data range between $v_c =
\SI{1e3}{cm\per s}$ and $\SI{4e3}{cm\per s}$.

\begin{figure}[ht]
    \centering
        \includegraphics[scale=0.9]{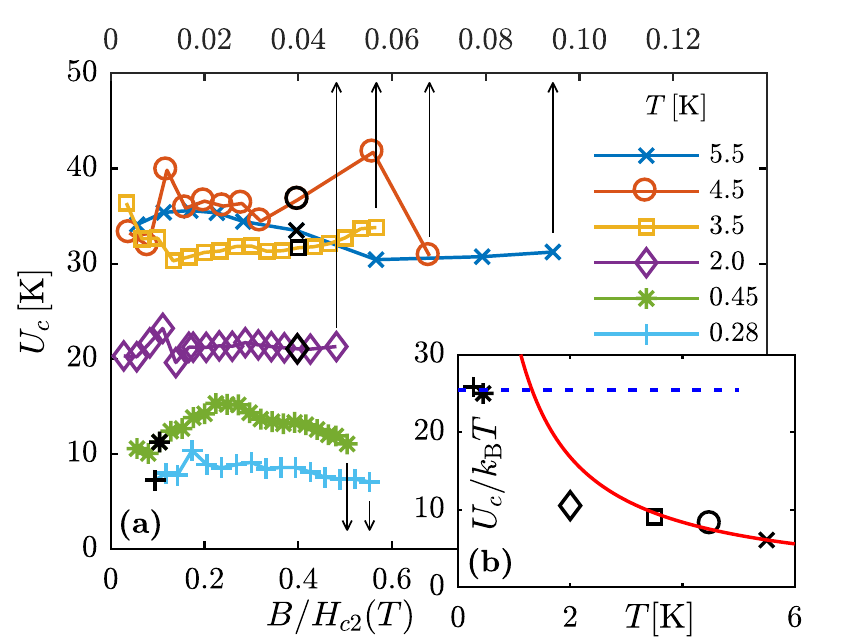}
	\caption{(a) Activation barriers $U_c$ in \textit{a}-MoGe versus
	magnetic field strength; lower and upper field axes refer to low ($T =
	\SI{0.28}{K}$ and $\SI{0.45}{K}$) and high temperature data. The
	decrease of $U_c$ with decreasing temperature $T$ can be explained in
	terms of a crossover to quantum creep, see inset. (b) Plotting the
	dimensionless thermal creep parameter $U_c/T$ versus temperature (see
	black markers; we chose similar values of the reduced field), one
	observes a crossover to a constant value at low temperatures, that is
	consistent with a saturation of $U_c/T$ (solid red line for a constant
	$U_c \approx \SI{34}{K}$) at the dimensionless action $S_c/\hbar$
	quantifying quantum creep (dashed blue line for a constant action
	$S_c$) when the latter takes over at small temperatures.}
	\label{Fig:Pratap_Uc}
\end{figure}

In a second step, we focus on the transition region of the rescaled data of
Fig.\ \ref{Fig:fits_Pratap}(a).  While the curves at higher temperatures are
rounded and shifted away from the excess-current characteristic, the data
taken at the two lowest temperatures collapse to an almost identical curve
after rescaling; note that the reduced temperatures $\tau_b$ are nearly equal
for the two curves, $\tau_b \approx 0.87$ versus $\tau_b \approx 0.86$ for the
curves 1 and 2, while the temperature $T$, quantifying thermal fluctuations,
increases by a factor $\approx 1.6$. Such a finding implies, that the voltage
response of the superconductor does not depend on temperature any more,
suggesting that quantum creep \cite{Blatter1991} may take over at these low
temperatures.  This hypothesis is further supported by comparing
the creep parameter $U_c/k_{\rm \scriptscriptstyle B} T$ extracted from fitting
the data for various temperatures, see Fig.\ \ref{Fig:Pratap_Uc}(b).
Their variations for temperatures above $\SI{3.5}{K}$ are consistent with a
value $U_c \approx 30 - \SI{40}{K}$, see \ref{Fig:Pratap_Uc}(a).
Extrapolating the ratio $U_c/k_{\rm \scriptscriptstyle B} T$ to the
low-temperature region results in values larger then observed, and hence much
lower vortex velocities.  This suggests that the thermal creep parameter
$U_c/k_{\rm \scriptscriptstyle B} T$ saturates at its quantum analog
$S/\hbar$, see Fig.\ \ref{Fig:Pratap_Uc}(b); the latter produces a still
appreciable (quantum) creep velocity $v \propto e^{-S/\hbar}$, with $S/\hbar <
U_c/k_{\rm \scriptscriptstyle B} T$.

\section{Activation barriers} \label{sec:barriers}
On a phenomenological level, creep-type vortex motion is a thermal process
with vortices escaping from defects by overcoming a drive-dependent activation
barrier $U(j)$; vortex motion then follows an Arrhenius-type formula for the
velocity $v = v_\th e^{-U(j)/k_{\rm \scriptscriptstyle B}T}$.  In comparison,
our equation \eqref{Eq:j-v} for the current--velocity characteristic
describes creep-type motion as well, but follows from a quantitative
determination \cite{Buchacek2018,Buchacek2019_theory} of the pinning force
density $F_\pin(v,T)$ entering the force-balance equation
\eqref{Eq:force_balance}.  In the following, we show how these
phenomenological and microscopic approaches relate to one another within
strong pinning theory. 

The interest in the activation barrier $U(j)$ is usually focused onto two
limits: i) at weak drives $j\to 0$, barriers either remain finite or diverge,
characterizing a vortex-liquid or a vortex-glass state, respectively
\cite{Blatter1994}, and ii) at drives $j \to j_c$, the barriers are expected
to vanish, $U(j) \approx U_c(1-j/j_c)^\alpha$, with an exponent $\alpha$
depending on the pinning model.  Strong pinning theory predicts
\cite{Buchacek2018, Buchacek2019_theory} a saturating barrier and a thermally
assisted flux-flow response at $j \to 0$. In the region near the critical
drive $j_c$ relevant in this study, the barriers vanish with an exponent
$\alpha = 3/2$; this result is universal for any smooth pinning potential.
However, strong pinning theory provides us with the further insight that the
appropriate variable in the barrier's scaling law is the pinning-force density
$F_\pin(v,T)$ rather then the Lorentz-force density $F_{\rm \scriptscriptstyle
L}$, see Eq.\ \eqref{Eq:U_F}.

The standard result with the exponent $\alpha = 3/2$ is straightforwardly
derived from our microscopic description in the limit where the viscous force
$-\eta v$ in Eq.\ \eqref{Eq:force_balance} (or the term $v/v_c$ in Eq.\
\eqref{Eq:j-v}) can be neglected.  In this situation, the characteristic
\eqref{Eq:j-v} is equivalent to the Arrhenius law with a barrier exponent
$\alpha = 3/2$: indeed, within this approximation, the driving force $F_{\rm
\scriptscriptstyle L}/F_c = j/j_c$ is balanced by the pinning force
$F_\pin(v,T)/F_c = [(k_{\rm \scriptscriptstyle B} T / U_c) \ln(v_\mathrm{th} /
v)]^{3/2}$ and our $v$--$j$ characteristic \eqref{Eq:j-v} can be cast into an
Arrhenius-law with a barrier $U(j) = U_c(1-j/j_c)^{3/2}$.

Upon increasing the drive $j$, however, the vortex velocity $v$ becomes
significant and the viscous term $-\eta v$ can no longer be neglected.
Combining Eq.\ \eqref{Eq:U_F} and the equation of motion
\eqref{Eq:force_balance}, we replace $F_\pin = F_{\rm \scriptscriptstyle L} -
\eta v$ to obtain a barrier that depends on both drive $j$ and velocity $v$ in
the form
\begin{align}
   U[F_\pin(v,T)] &= U_c\Bigl(1-\frac{F_{\rm \scriptscriptstyle L}-\eta v}{F_c}\Bigr)^{3/2}\\
   \label{Eq:barrier_modified}
   &= U_c \Bigl[1 - \Bigl(\frac{j}{j_c} - \frac{v}{v_c}\Bigr)\Bigr]^{3/2} 
   \equiv U(j,v).
\end{align}
Using this expression for the barrier, the characteristic given by Eq.\
\eqref{Eq:j-v} can be written as a \textit{self-consistent} Arrhenius law for
the vortex velocity $v(j)$,
\begin{align}\label{Eq:modified_Arrhenius}
   v = v_\th e^{-U(j,v)/k_{\rm \scriptscriptstyle B} T}.
\end{align}
\begin{figure}[b]
    \centering
        \includegraphics[scale=1]{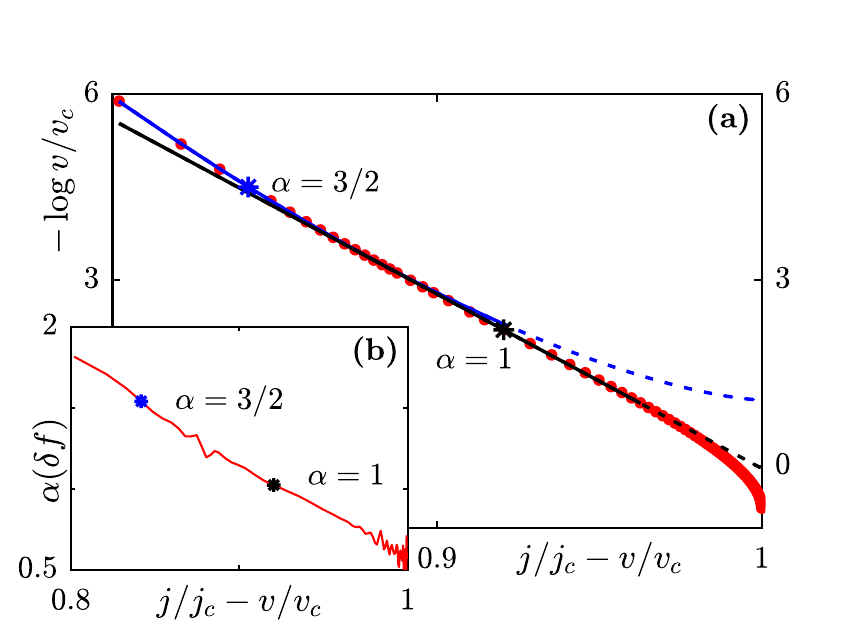}
	\caption{Analysis of activation barriers for 2H-NbSe$_2$, with data
	for $B = \SI{1.0}{T}$, $T = \SI{5.2}{K}$ represented by red dots.
	Shown are the $j$--$v$ data represented as $-\log(v/v_c) =
	U/k_{\rm\scriptscriptstyle B}T + \gamma$ versus $j/j_c - v/v_c =
	F_\pin/F_c$. The activation barrier $U(F_\pin)$ in (a) is fitted to
	$U(F_\pin) = U_c (1- F_\pin/ F_c)^\alpha$ for different exponents
	$\alpha = 3/2$ (blue) and $\alpha = 1$ (black).  The fits outside the
	range of validity $v < v_\th/e$ are continued with dashed curves. The
	inset (b) shows the construction of points $\alpha = 1$ and $\alpha =
	3/2$ around which the fitting is done.  \label{Fig:Andrei_U}}
\end{figure}

Alternatively, using the data of the scaled $j$--$v$ characteristic, the
expression \eqref{Eq:barrier_modified} provides us with a direct access to the
activation barrier $U(j,v)$. Indeed, plotting $\log(v_c/v)$ versus
$j/j_c - v/v_c$, we represent $U[F_\pin(v,T)]/k_{\rm
\scriptscriptstyle B} T + \gamma$ versus $F_\pin(v,T)/F_c$ with the offset
$\gamma = \log(v_c/v_\th)$.  Fig.\ \ref{Fig:Andrei_U} shows a typical outcome
of arranging the data in this new manner.  We then can follow two strategies,
i) either assume the validity of \eqref{Eq:barrier_modified} and use this
fitting ansatz to extract the creep parameters $U_c/k_{\rm \scriptscriptstyle
B} T$ and $v_\th/v_c$, or ii) interpret the data as directly providing the
functional form of $U[F_\pin(v,T)]$, up to a constant.

The parameters $U_c$ and $v_\th/v_c$ shown in Figs.\
\ref{Fig:parameters_Andrei} (c) and (d) and in Fig.\ \ref{Fig:Pratap_Uc}(a)
have been obtained by following the procedure i).  In extracting the parameter
$U_c/k_{\rm \scriptscriptstyle B} T$, we have to select the appropriate
portion of the curve: starting at $j/j_c-v/v_c =  1$ (corresponding to
$F_\pin/F_c = 1$) in Fig.\ \ref{Fig:Andrei_U}, the barrier $U(F_\pin)$
initially grows with a concave shape, goes through an inflection point, and
then continues in a convex curve at smaller values $F_\pin/F_c  < 0.9$.  The
initial concave form for $F_\pin$ close to $F_c$ originates from the
saturation of $F_\pin(v,T) \to F_c$ when the velocity $v$ increases beyond the
thermal velocity $v_\th$. The creep-dominated region at small velocities
corresponds to the convex region in Fig.\ \ref{Fig:Andrei_U} and it is this
region that provides us with the value for the reduced barrier $U_c/k_{\rm
\scriptscriptstyle B} T$.  The ratio $v_\th/v_c$ derives from the condition
$U[F_\pin(v,T)/F_c = 1]=0$, i.e., the offset $\gamma$ in $\log(v_c/v)$ at
$j/j_c - v/v_c = 1$, once the curve $U[F_\pin(v,T)]$ has been fitted and
extrapolated to $F_\pin/F_c = 1$.

In a systematic fit, we search for the region that is best described by Eq.\
\eqref{Eq:barrier_modified}. We define the rescaled pinning force $\delta f =
1-F_\pin/F_c$ and take the derivative $\partial_{\delta f} \log(v_c/v)$ in
order to eliminate the constant shift $\gamma$; taking the log-derivative, we
obtain $\alpha = 1 + \partial \log[\partial_{\delta f} \log(v_c/v)]/\partial
\log(\delta f)$ as shown in the inset Fig.\ \ref{Fig:Andrei_U}(b)).  The fit
to Eq.\ \eqref{Eq:j-v} then is done around the point $\delta f^*$ for which
$\alpha(\delta f^*) = 3/2$. 

In following the alternative procedure ii) instead, we directly obtain the
shape of $U(F_\pin)$ (up to a constant shift) but miss its analytic form.
Furthermore, we have to be careful in interpreting the data as a barrier, as
the latter requires the quantity $U_c/k_{\rm \scriptscriptstyle B} T$ to be
large. In an attempt to extract some effective functional form, we can make
some more progress by using our findings for the exponent $\alpha(\delta f)$,
which, as shown in Fig.\ \ref{Fig:Andrei_U}(b), is not at all a simple
constant. For instance, it is possible to find a region at higher currents in
Fig.\ \ref{Fig:Andrei_U}(a) where the barrier shape is better characterized by
an exponent $\alpha = 1$. Such a linear dependence of the activation barrier
reminds about the original assertion by Anderson \cite{Anderson1962} for the
creep barrier as emerging from the competition between the defect's pinning
energy $\sim H_c^2 d^3/8\pi$ and the Lorentz force energy $j B d^4/c$ of flux
bundles with volume $d^3$; the corresponding creep barrier then can be written
in the simple form $U(j) = U_c (1-j/j_c)$. In Fig.\ \ref{Fig:Andrei_U}(a), we
compare separate fits to the data with $\alpha = 3/2$ (blue line) and $\alpha
= 1$ (black line).  At first sight, the fit for $\alpha = 1$ looks rather
good, in particular at higher drives. Indeed, the changeover from a convex to
a concave form at large pinning forces $F_\pin$ produces an inflection point
with a region where the exponent $\alpha = 1$ quite naturally provides a
better match to the data.  However, this region close to $j_c$ is flow- rather
than creep-dominated, with a barrier of order or even smaller than $k_{\rm
\scriptscriptstyle B} T$. It then is the region at smaller drives and
velocities where creep effects are expected to manifest in a clean and
unperturbed manner. At these smaller drives, it is the exponent $\alpha = 3/2$
that provides a consistent description of the experimental data. Going to even
smaller drives, our expansion $U \propto (\delta f)^{3/2}$ is expected to
break down \cite{Buchacek2019_theory} and the shape of $U(F_\pin)$ depends on
the detailed form of the pinning potential.

\begin{figure}[b]
\centering
\includegraphics[scale=1]{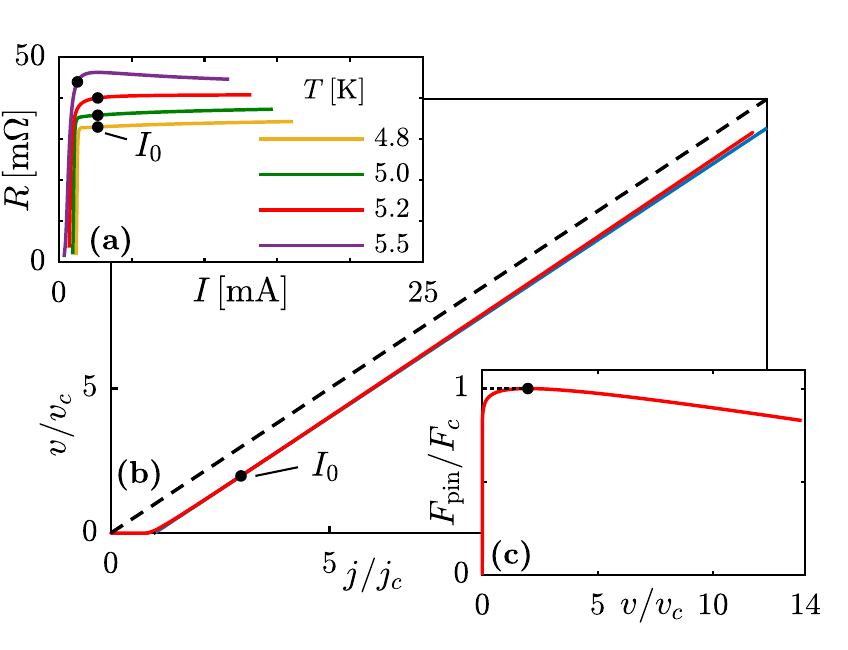}
\caption{Data for 2H-NbSe$_2$ ($B = \SI{1.0}{T}$) covering the complete measured range of
currents.  (a) The differential resistance $R = dV/dI$ exhibits a sharp
increase and then becomes flat near the flux-flow resistance $R_\mathrm{ff}$;
for the case of $T = \SI{5.5}{K}$, the resistance goes through a maximum at
low drives, while we expect a broad maximum at larger drives for the other
temperatures. The black dot marks the current $I_0$ chosen for the linear
extrapolation.  (b) Current--velocity characteristic (red for $T =
\SI{5.2}{K}$) after rescaling the experimental data.  The excess-current
characteristic (blue straight line) is constructed by linear extrapolation
around the current $I_0$. At large currents, the creep characteristic (red
line) approaches the free flux-flow (dashed). (c) The pinning-force density
$F_\pin$ (red for $T = \SI{5.2}{K}$) rises steeply towards the maximum value
$F_c$ near the velocity $v_\th$ and then slowly decays, in agreement with the
theoretical prediction \cite{Thomann2017}.} \label{Fig:Andrei_rho}
\end{figure}

Finally, we comment on some subtleties in using the high velocity data for the
extraction of $v_c$ (via $\rho_\mathrm{ff}$) and $j_c$.  The strong pinning
theory provides a detailed picture of vortex motion on several velocity
scales:  thermal creep \cite{Buchacek2018, Buchacek2019_theory} as discussed
above affects the vortex response only at velocities below $v_\th$. Beyond
$v_\th$, creep effects eventually become irrelevant and the finite temperature
creep- and $T=0$ excess-current characteristics merge. This merging in a
shifted linear flux-flow characteristic manifests in the data as a saturation
of the differential resistance, see Fig.~\ref{Fig:Andrei_rho}(a). 

In performing accurate fits, it is important to have experimental data that
goes beyond the merger at $v_\th$ and reaches some saturation in the
differential resistance, as this part of the data allows for inferring the
correct critical current density $j_c$ and flux-flow resistivity
$\rho_\mathrm{ff}$ through linear extrapolation. In an ideal situation, the
linear excess-current characteristic above $v_\th$ allows for a
straightforward extrapolation. In reality, the characteristic is never
perfectly linear and the extrapolation has to be done around some chosen point
$I_0,V_0$ of the characteristic. In order to find the correct point $I_0$
around which to construct the linear extrapolation, we combine the creep
dynamics discussed here with the results of the flow-dynamics described in
Refs.\ [\onlinecite{Thomann2012,Thomann2017}].

\begin{figure}[b]
\centering
\includegraphics[scale=1]{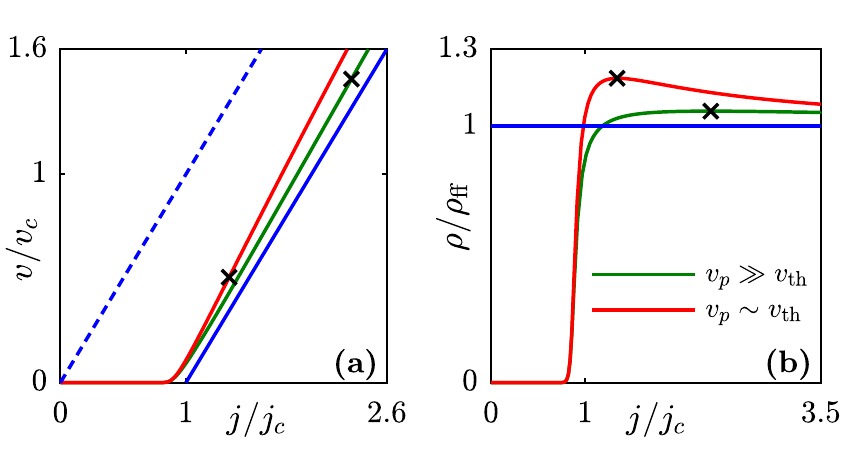}
     \caption{Illustrative sketch of current--voltage characteristic (left,
     (a)) and differential resistance (right, (b)) for the two cases with
     $v_\th \ll v_p$ (in green) and $v_p \sim v_\th$ (in red); blue dashed
     (solid) lines are (shifted) flux-flow curves. The separation of scales is
     a necessary condition for the appearance of the straight excess-current
     characteristic in (a), see green curve. The scaled differential
     resistivity shown in (b) rises steeply on the thermal creep scale
     $v_\mathrm{th}$, overshoots the flux-flow resistance $R_\mathrm{ff}$, and
     then smoothly approaches $R_\mathrm{ff}$ from above. The resulting
     maximum (black cross) is broad when velocity scales are separated, $v_\th
     \ll v_p$ (in green); this reminds about the behavior seen in Fig.\
     \ref{Fig:Andrei_rho} for the lowest three temperatures. On the other
     hand, a more narrow maximum appears near $v_\th$ when $v_p \sim v_\th$
     (in red), that resembles more the behavior of the high-temperature data
     at $T = \SI{5.5}{K}$ in the experiment of Fig.\ \ref{Fig:Andrei_rho}.}
     \label{Fig:res}
\end{figure}

The generic strong pinning situation is characterized by a separation of
scales $v_c \ll v_p$, where $v_p \sim f_p/\eta a_0^3$ is the velocity scale
for dissipative vortex motion within the defect potential excerting typical
forces $f_p$. The excess current--voltage curve then is expected to turn back
towards {\it free} flux-flow (see dashed blue line in Fig.\ \ref{Fig:res}(a)) only
at velocities larger than $v_p$.  Provided that $v_\th \ll v_p$, one thus
expects the differential resistance to rise sharply towards $R_\mathrm{ff}$ on
the scale $v_\th$, then going through a broad maximum within the extended
velocity region $v_\th < v < v_p$, and returning from above to the free
flux-flow branch and hence to $R_\mathrm{ff}$ at very large currents $v \gg
v_p$, see the sketch (green line) in Fig.\ \ref{Fig:res}(b).  This behavior is
well in line with the behavior of the characteristics in Fig.\
\ref{Fig:Andrei_rho} at the three lower temperatures.  For the data measured
at $T = \SI{5.5}{K}$ (magenta line in Fig.\ \ref{Fig:Andrei_rho}(a)) the
maximum is sharper and appears at lower drives.  This is consistent with the
scenario shown in Fig.\ \ref{Fig:res}(b), red line, where $v_p$ is of the
order of $v_\th$.  In this case, the differential resistance overshoots
$R_\mathrm{ff}$ and goes through a more narrow maximum close to $v_\th$.

In choosing the point $I_0$ where the extrapolation to the excess-current
characteristic (blue line in  Fig.\ \ref{Fig:Andrei_rho}(b)) should be done,
we should stay below the maximum in $R$ (i.e., below the inflection point of
the current--voltage characteristic), ideally at the crossing of the
differential resistance $R$ with the flux-flow resistance $R_\mathrm{ff}$. In
particular, this educated choice guarantees that the extra\-polation never
crosses the $I$--$V$ characteristic but rather touches the latter in a
tangent at $I_0$. Unfortunately, the flux-flow resistance $R_\mathrm{ff}$
may not be accurately known, which leaves some arbitraryness in the choice of
$I_0$. For the lowest three temperatures in Fig.\ \ref{Fig:Andrei_rho}(a), we
have chosen a value $I_0 = \SI{2.7}{mA}$ near the onset of the flat maximum in
$R$ and remark that the extracted parameters do not differ significantly for
somewhat different choices of $I_0$. For the highest temperature $T =
\SI{5.5}{K}$, we have chosen a value $I_0 = \SI{1.3}{mA}$ before the maximum
(which is expected at values $R > R_\mathrm{ff}$); again, the precise choice
of $I_0$ does not change the extracted pinning parameters in a significant
manner.  Note that the experimental access to such a high-velocity regime is
quite problematic in general due to heating effects that may even destroy the
sample; this type of analysis then is restricted to materials with a small
ratio $j_c/j_0$.

The above discussion also sheds some more light on the concave region of the
barrier plot $U(F_\pin)$ in Fig.\ \ref{Fig:Andrei_U} close to $F_c$.
Translating the behavior of the current--voltage and resistance curves in
Fig.\ \ref{Fig:res} to the pinning-force density $F_\pin(v)$, one notes that
the latter exhibits a broad maximum or plateau $F_\pin(v_\th < v < v_p)
\approx F_c$ for the case $v_p \gg v_\mathrm{th}$, while a more narrow maximum
around $v_\th$ is expected when $v_p \sim v_\th$.  Indeed, the maximum in
$F_\pin$ corresponds to the point $\rho/\rho_\mathrm{ff} = 1$ in Fig.\
\ref{Fig:res}(b) and the derivative $\partial_j \rho$, small (large) for the
case $v_p \gg v_\mathrm{th}$ ($v_p \sim v_\th$), determines the curvature of
$F_\pin$.  In Fig.\ \ref{Fig:Andrei_rho}(c), we show the pinning-force density
(with a broad maximum) extracted from the data at $T = \SI{5.2}{K}$. We then
can expand $F_\pin(v)$ around its maximum and approximate $\delta f \propto
[(v_0 - v)/v_c]^2$.  Inverting and expanding for small $\delta f \ll 1$ 
provides us with the scaling $\log(v_0/v_c) - \log(v/v_c) \propto \delta
f^{1/2}$.  The appearance of the concave region in $U(F_\pin)$ with an
exponent $\alpha \approx 1/2$ is thus the direct consequence of the maximum in
$F_\pin$ at $F_c$.

\section{Parameters from strong pinning theory}\label{sec:par}

The interaction of vortex lines with sufficiently strong defects gives rise to
bistable solutions for the vortex lattice displacement; the appearance of such
bistabilities is the hallmark of the strong pinning regime. The weak- to
strong pinning crossover is characterized by the Labusch parameter
\cite{Labusch1969, Blatter2008} $\kappa = \max [-e_p''(r)]/\C$, comparing the
maximum (negative) curvature of the defect pinning potential $e_p(r)$ near its
edge with the effective vortex lattice stiffness $\bar{C}$; pinning is strong
provided that $\kappa > 1$. A second condition on the applicability of strong
pinning theory is the independent action of individual pins, requiring that
the density $n_p$ of pins is small, $n_p \xi^2 \kappa a_0 < 1$. Therefore,
strong pinning does not necessarily imply a large critical current density
$j_c$. As the density $n_p$ is increased above (or the field $B =
\Phi_0/a_0^2$ decreased below) this condition, 3D strong pinning goes over
into 1D strong pinning of individual vortices \cite{Blatter2004}. Below, we
cite the main results of the 3D strong pinning regime as relevant in the
present discussion and show how pertinent strong-pinning parameters such as
$n_p$, $\kappa$, $f_p$, $U_c$ can be extracted from the comparison of
theoretical predictions with experimental data, at least in principle.

Vortices (with a core of size $\xi$) remain pinned on a defect over an area
$S_\trap \approx t_\parallel t_\perp$, with $t_\parallel \approx \kappa \xi$
and $t_\perp \approx \xi$ the longitudinal (along the vortex motion) and
transverse trapping lengths.  Assuming each defect to exert a pinning force
$f_p\sim e_p/\xi$ ($e_p$ is the pinning potential depth) on the vortex, we
find the maximal (or critical) pinning-force density $F_c = n_p
(S_\trap/a_0^2)f_p$.  When approaching the boundary of strong pinning at
$\kappa \to 1$, the pinning-force density is reduced by a factor
$(\kappa-1)^2$ and we can make use of the interpolation formula
\begin{equation} \label{eq:F_c}
   F_c\approx \gamma n_p(\xi^2 \kappa /a_0^2)(e_p/\xi)(1-1/\kappa)^2.
\end{equation}
The numerical $\gamma$ can be calculated once the specific shape of the
pinning potential is known \cite{Buchacek2018}; for the Lorentzian pinning
potential $e_p(r) = e_p/[1+(r^2/2\xi^2)]$, we find that $\gamma\approx 0.4$.

The intrinsic field- and temperature dependence of the critical current $j_c=
cF_c/B$ follows from the corresponding dependencies of $e_p$, $\kappa$, and
$\xi$. In the vicinity of the upper-critical field $H_{c2}(T)\approx
H_{c2}(0)\, \tau$ with $\tau = 1- T/T_c$, the coherence- and London
penetration lengths scale as $\xi = \xi_0\tau^{-1/2}$ and $\lambda =
\lambda_0\tau_b^{-1/2}$, respectively, with $\tau_b = \tau - b$ and $b =
B/H_{c2}(0)$.  Various pinning models involving metallic and insulating
defects or $\delta T_c$-pinning, have been discussed in Ref.\
[\onlinecite{Willa2016}]; the pinning potential depth $e_p$ and the pinning
strength $\kappa$ then depend in various ways on $\lambda$ and $\xi$. It turns
out that the dominant contribution to the scaling near the $H_{c2}(T)$-line
appears through the pinning energy $e_p = e_{p0} (1-t-b)^{\beta_e}$ and
the Labusch parameter $\kappa = \kappa_0 (1-t-b)^{\beta_\kappa}$ with
model-dependent exponents $\beta_e$ and $\beta_\kappa$.

Scaling the critical current density $j_c = cF_c/B$ with the
(zero-temperature) depairing current density $j_0 = (2/3\sqrt{3})c
H_{c}^2(0)\xi_0/\Phi_0$ (with $H_c(T)$ the thermodynamic critical field), we
find that this ratio only involves the effective defect number in the trapping
volume $n_p \xi^2 \kappa a_0 = n_p S_\trap a_0 < 1$ and the ratio $e_p/e_0
\lesssim 1$ with $e_0 = H_c^2(0) \xi_0^3 /8\pi$ the (zero-temperature)
condensation energy,
\begin{align} \label{Eq:j_c}
   \frac{j_c}{j_0} \approx \frac{3\sqrt{3}\gamma}{16\pi}
   \, n_p \xi^2 \kappa a_0\, \frac{e_p}{e_0} \frac{\xi}{a_0}\, 
   (1-1/\kappa)^2 (1-T/T_c)^{1/2}.
\end{align}
With $\kappa \propto a_0$, we find the typical strong-pinning scaling
\cite{IvlevOvchinnikov91,Thomann2017,Kwok2016} $j_c \propto 1/\sqrt{B}$. Upon
decreasing the field below the 3D strong pinning condition $n_p S_\trap a_0 <
1$, pinning turns one-dimensional (1D) and $j_c$ is expected to saturate to a
$B$-independent value.  Taking into account a weak $\kappa$-dependence of the
transverse trapping $t_\perp\sim \kappa^{1/4}\xi$ \cite{Thomann2017} changes
the field-scaling of the critical current density to $j_c\propto B^{-\alpha}$
with $\alpha = 5/8$ for a Lorentzian-shaped pinning potential.  This result
has been verified and augmented by numerical simulations \cite{Willa2017}
showing that the exponent $\alpha$ in fact decreases for increasing defect
densities or vortex core size.

The result \eqref{Eq:j_c} tells, that $j_c$ should decrease on approaching the
$H_{c2}(T)$-line, in agreement with the finding in Fig.\
\ref{Fig:parameters_Andrei}(a).  The scaling $j_c \propto B^{-\alpha}$ with
exponents $\alpha \lesssim 0.5$ is observed in the data of Fig.\
\ref{Fig:Pratap_j_c} (a) and (b).

Next, we discuss the scale $U_c$ of the activation barrier. This turns out
proportional to the pinning energy $e_p$, vanishes on approaching weak pinning
$\kappa \to 1$, and only weakly depends on $\kappa$ for very strong pinning
\cite{Buchacek2018}; it is accurately described by the interpolation formula
\begin{align}\label{Eq:U_c}
  U_c \approx \tilde{g} e_p\, (1-1/\kappa)^2
\end{align}
with the numerical $\tilde{g}\approx 0.4$ for Lorentzian pinning potential.
While the values of $U_c$ in Fig.\ \ref{Fig:parameters_Andrei}(c) decrease
rapidly on approaching the $H_{c2}(T)$-line, the data in
Fig.~\ref{Fig:Pratap_Uc} is more consistent with a constant value. Indeed,
substantial variations of the barrier with field and temperature are to be
observed only sufficiently close to the $H_{c2}(T)$-line; this is the case for
2H-NbSe$_2$ where $\tau_b < 0.16$. On the other hand, the data on
\textit{a}-MoGe has been obtained further away from the $H_{c2}$-line, with
values of $\tau_b$ all larger than 0.2.

Comparing Eq.~\eqref{Eq:j_c} for $j_c$ with the expression \eqref{Eq:U_c} for
the activation barrier $U_c$ allows us to extract the effective defect number
from the experimental data,
\begin{align}
\label{Eq:n_p}
   n_p \xi^2\kappa a_0 \approx
   \frac{16\pi \tilde{g}}{3\sqrt{3}\gamma}\frac{j_c}{j_0}
   \frac{e_{0}}{U_c} \frac{a_0}{\xi} \frac{1}{(1-T/T_c)^{1/2}}.
\end{align}
Given the values of $j_c$ and $U_c$, we can use Eq.\ \eqref{Eq:n_p} to find an
estimate for the defect parameter $\kappa n_p a_0 \xi^2$; the data on
2H-NbSe$_2$ provides us with the values $\kappa n_p a_0 \xi^2 \approx
(1.3,\,1.7,\,3.4,\,12) \times 10^{-4}\ll 1$ at the four different
temperatures, all consistent with the assumption of 3D strong pinning.  With
$\xi_0 \approx \SI{77}{\AA}$ \cite{Banerjee1997} and assuming a value of order
unity for $\kappa$ provides the estimate $n_p \sim \SI{2e15}{cm^{-3}}$.
\begin{figure}[b]
\centering
\includegraphics[scale=1]{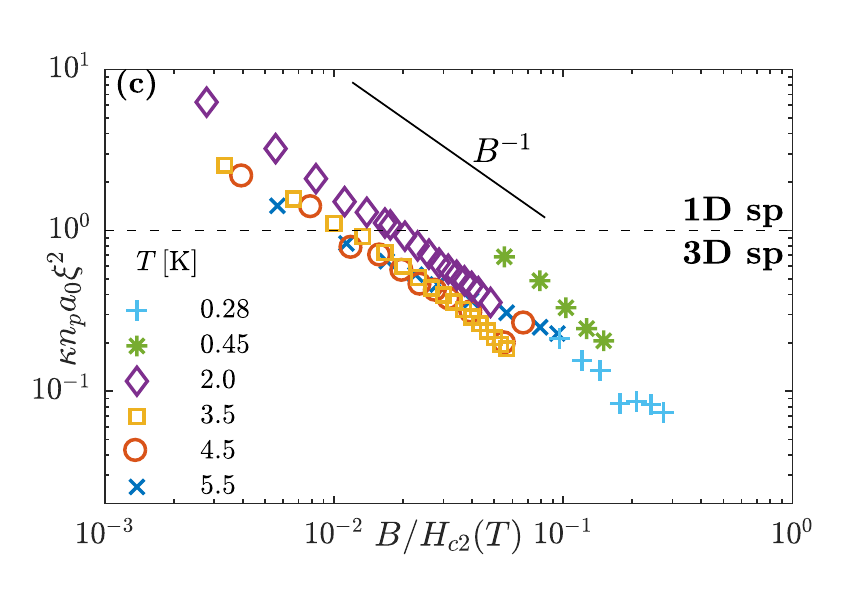}
     \caption{Defect parameter $n_p \xi^2 \kappa a_0$ as a function of
     magnetic field scaling with $\propto B^{-1}$, as expected from strong
     pinning theory.  The dashed line marks the rough position of the expected
     crossover line between the 1D- and 3D-strong pinning regimes in the
     pinning diagram of Ref.\ [\onlinecite{Blatter2004}].}
\label{Fig:Pratap_n_p}
\end{figure}

The density parameter $n_p \xi^2 \kappa a_0$ extracted from critical currents
$j_c$ and activation barriers $U_c$ as derived from the fits on the $a$-MoGe
films, see Figs.\ \ref{Fig:Pratap_j_c} and \ref{Fig:Pratap_Uc}, is shown as a
function of field in Fig.\ \ref{Fig:Pratap_n_p}. At large fields $B > 4$ T,
the separation between vortices $a_0$ is smaller than the film thickness $d$,
the density parameter $\kappa n_p a_0 \xi^2 < 1$ is small, and we expect 3D
strong pinning.  On decreasing the field, two things happen: i) as the
distance $a_0$ bet\-ween vortices (that equals the extent of the distortion
along pinned vortices) drops below $d$, vortices are cut and we enter the 2D
strong pinning regime that is still well described by our strong pinning
theory but with a modified effective elasticity $\bar{C} \propto B$ involving
only shear. As a result, the Labusch parameter scales as $\kappa \propto
a_0^2$ and the critical current density $j_c \propto 1/B$. ii) With increasing
density parameter $n_p \xi^2 \kappa a_0$, vortices become individually pinned,
either as 1D lines (at high fields with $a_0 < d$) or as 0D Pearl vortices (at
low fields with $a_0 > d$).  In this case, the critical current density $j_c$
is expected to flatten and become independent of field $B$. The critical
current density $j_c$ in Fig.\ \ref{Fig:Pratap_j_c} seems to flatten at the
lowest fields (see data at $T = \SI{3.5}{K}$ and $T = \SI{4.5}{K}$) that may
indicate a crossover to a field independent 1D or 0D regime. Furthermore, the
field scaling $j_c \propto B^{-\alpha}$ with an observed $\alpha$ between 0
and unity covers the range of expected behavior, however, without clear
attribution to a specific regime.  An accurate association with a specific
pinning region then seems difficult in the low-field/high-density region,
given the competition between the dimensional crossover and the density $n_p$
crossover.  Finally, we can use the data to extract an estimate for the defect
density: with $\xi_0\approx \SI{52}{\AA}$ and $\kappa$ of order unity, the
defect density $n_p$ itself assumes a value of order $n_p \sim
\SI{1e17}{cm^{-3}}$.
\begin{figure}[h]
\centering
\includegraphics[scale=1]{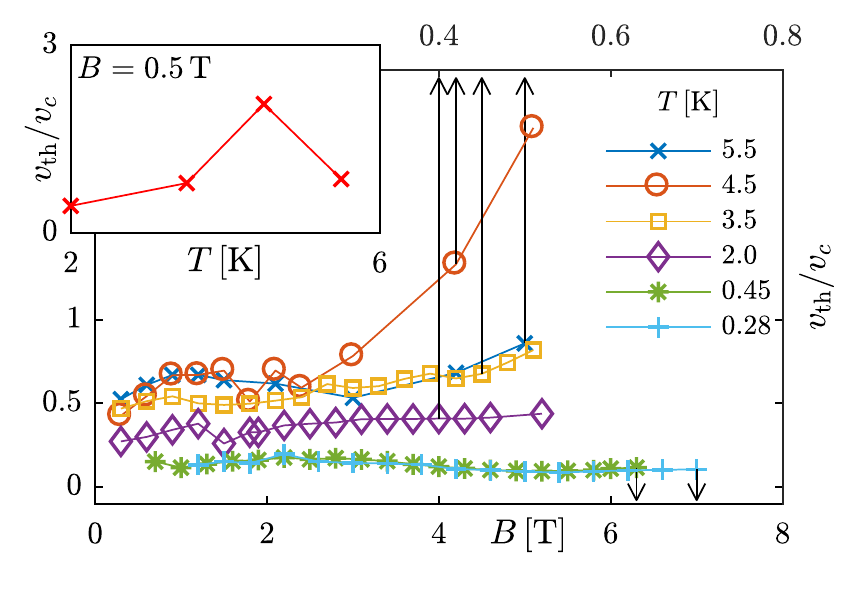}
     \caption{Scaled thermal velocity $v_\th/v_c$ versus magnetic field $B$;
     upper and lower field axes refer to low ($T = \SI{0.28}{K}$
     and $\SI{0.45}{K}$) and high temperature data.
     The inset shows the dependence on temperature at a constant field $H = \SI{0.5}{T}$,
     with a non-monotonic dependence as found before in the data for 2H-NbSe$_2$,
     see Fig.\ \ref{Fig:parameters_Andrei}(d).
     }
\label{Fig:Pratap_v_th}
\end{figure}

Last, we turn our interest to the thermal velocity parameter $v_\th/v_c$.
The theoretical prediction \cite{Buchacek2018} 
\begin{align}\label{Eq:v_th_0}
   \frac{v_\th}{v_c} \approx \frac{T}{e_p} \frac{a(\kappa)}{n_p a_0\xi^2}
\end{align}
is based on a simple particle-like ansatz in Kramer's rate expression
\cite{Kramers1940,Hanggi1990,Buchacek2019_theory}; its proper evaluation, both
theoretically and from experiment is notoriously difficult as it appears as
the prefactor in the thermal activation rate which is dominated by the
exponential factor with its activation barrier.  Approximating $a(\kappa)
\approx \tilde{a} \, (1-1/\kappa)^{-3/2}$ with a typical value $\tilde{a}
\approx 0.1$, and using Eq.\ \eqref{Eq:U_c}, we arrive at the simpler result
\begin{align}\label{Eq:v_th}
   \frac{v_\th}{v_c} \approx \biggl[\frac{k_{\rm\scriptscriptstyle B}T}{U_c}
   \frac{\tilde{a}\tilde{g}}{n_p \xi^2 \kappa a_0}\biggr]\, [\kappa\, (1-1/\kappa)^{1/2}],
\end{align}
where we have suitably factorized the result for later convenience, see below.
This result predicts an increase of ${v_\th}/{v_c}$ with temperature (due to
the factor ${k_{\rm\scriptscriptstyle B}T}/{U_c}$) that is consistent with the
findings obtained from fitting the characteristic, see Figs.\
\ref{Fig:parameters_Andrei}(d) and \ref{Fig:Pratap_v_th}, apart from the
datapoint at the highest temperature. This might be explained by a collapse of
the factor $(1-1/\kappa)^{1/2}$ near the $H_{c2}$-line that occurs in several
of the pinning models discussed in Ref.\ [\onlinecite{Willa2016}]. 

Alternatively, the first factor of Eq.\ \eqref{Eq:v_th} can be evaluated using
the experimental findings for $U_c$ and values for $n_p\xi^2\kappa a_0$ from
Eq.\ \eqref{Eq:n_p} (derived from experimental results for both $U_c$ and
$j_c$); such an analysis provides a result of order unity for the first
factor, about an order of magnitude larger than the values extracted from the
characteristic shown in Fig.\ \ref{Fig:parameters_Andrei}(d). Consistency then
would require $\kappa$ to be close to unity, i.e., individual pins are
marginally strong. Repeating this analysis for $a$-MoGe and using known
experimental results for $U_c$ and $j_c$, we find a small value of order
$10^{-2}$ for the first factor in Eq.\ \eqref{Eq:v_th}. One then concludes
that the Labusch parameter $\kappa$ should be large in $a$-MoGe in order to
reach consistency with the results in Fig.\ \ref{Fig:Pratap_v_th}.  However, a
word of caution is in place here, as both our theoretical knowledge on the
preexponential factor $v_\th/v_c$ as well as our precision to extract a
reliable value from Fig.\ \ref{Fig:Andrei_U} are quite limited at this stage.

\section{Summary and conclusion}\label{sec:sc}

By applying the quantitative theory of strong pinning to current--voltage
measurements, we provided a first quantitative data-driven analysis of vortex
creep in the critical region and thus demonstrated the potential of the
strong-pinning paradigm for explaining pinning and creep in superconductors.
The strong pinning paradigm comes with a number of microscopic assumptions:
defects have to be strong, i.e, they must generate bistable pinning states, and
their density has to be small such that they act independently.

In return, we obtain specific phenomenological predictions: the critical
current density follows a field-scaling $j_c\propto B^{-\alpha}$ with
$\alpha\approx 0.5$ that is different from weak collective pinning theory and
pinning persists well-beyond the critical drive that results in a linear
excess-current characteristic. The experimental data analyzed in our work
satisfies these requirements and provides a coherent picture when submitted to
a strong pinning analysis.  Studying 2H-NbSe$_2$ and \textit{a}-MoGe with
moderate critical temperature, we demonstrated that high temperature is not a
necessary requirement for significant creep effects on the current--voltage
characteristic. Indeed, the sensitivity of the characteristic to thermal
fluctuations follows from the creep parameter $U_c/k_{\rm \scriptscriptstyle
B} T$ which becomes small near the upper-critical field; temperature- and
field variations of $U_c$ then have a large influence on the characteristic
and are visible through thermal creep effects.

The barriers $U_c$ extracted from the fits can be compared with experiments on
persistent current relaxation quantified by the normalized creep rate $S =
-\partial \log j/ \partial \log t$ \cite{Blatter1994}. Assuming that the
activation barrier $U(j)$ vanishes with a characteristic exponent $\alpha =
3/2$, the creep rate is related to the barrier through $S \approx
(2/3)(k_\mathrm{B}T/U_c)^{2/3}$ \cite{Buchacek2019_theory}.  Fitting the data
of 2H-NbSe$_2$ for $T = \SI{4.8}{K}$ yields the barrier $U_c\approx
\SI{980}{K}$, see Fig.\ \ref{Fig:parameters_Andrei}, a value that is
consistent with the observed creep rate \cite{Eley2018} ranging from $S
\approx 5\times 10^{-3}$ to $S\approx 10^{-2}$.

It is also important to stress that while the prefactor $U_c$ defines the
barrier scale due to the defect potential (and is comparable to the defect
pinning energy $e_p$), the actual barrier $U(F_\pin)$ relevant for creep is
much reduced due to the drive, what renders the creep motion visible in the
experiment. For large drives, this barrier eventually drops below the
fluctuation energy $k_{\rm \scriptscriptstyle B} T$ and Kramer's rate theory
breaks down. This restricts the applicability of our results and thus the
reliability of the fits to the region $v < v_\th/e$.  At large velocities $v >
v_p$ with $v_p \gg v_\th$, dynamical effects become important and experimental
data covering such a region far beyond the critical current then show a collapse
of the pinning force and an approach towards the free flux-flow, again in
agreement with the strong pinning theory.

Following the prediction of strong pinning theory that the activation barrier
$U(F_\pin)$ depends on the pinning-force density $F_\pin$ rather than the
driving current density $j$, we have proposed a new methodology to extract the
creep parameters $U_c/k_{\rm \scriptscriptstyle B} T$ and $v_\th/v_c$, the
barrier and prefactor in the Arrhenius law for the activated process.  In
comparison to the standard assumption of a barrier dependence $U(j/j_c)$, the
strong pinning expression includes the dissipative force as well,
$U(j/j_c-v/v_c)$; the two Ans\"atze coincide in the region of very small
velocities or large barriers, where the dissipative term can be ignored,
$v/v_c \ll j/j_c$.  Analyzing our data, it turns out that this correction is
relevant: e.g., for \textit{a}-MoGe and $T = \SI{3.5}{K}$
(Fig.~\ref{Fig:fits_Pratap}), we find that $v/v_c\approx 0.11$ for $j =
0.8j_c$; hence, neglecting the viscous term would lead to a shift $\delta j
\approx 0.11 j_c$ of the theoretical prediction and hence a significant
deviation from the experimental data.  Finally, the intriguing saturation of
the creep parameter $U_c/k_{\rm \scriptscriptstyle B} T$ in \textit{a}-MoGe at
low temperatures points to the possibility of performing a direct observation
of quantum creep through current--voltage measurements, that could be verified
by future experimental and theoretical work.

\acknowledgements
We thank Kristin Willa, Eli Zeldov, Marcin Konczykowski, and Roland Willa for
inspiring discussions. M.B.\ acknowledges financial support of the Swiss
National Science Foundation, Division II. Z.L.X.\ acknowledges supports by the
U.S. Department of Energy, Office of Science, Basic Energy Sciences, Materials
Sciences and Engineering and the National Science Foundation under Grant No.
DMR-1901843.

\bibliography{bib_exp_paper}

\end{document}